\documentclass[12pt]{article}

\usepackage{epsfig}
\usepackage{amssymb,amsmath}

\newcommand{\bea}{\begin{eqnarray}}
\newcommand{\eea}{\end{eqnarray}}

 \makeatletter
\def\alt{\mathrel{\mathpalette\gl@align<}}
\def\agt{\mathrel{\mathpalette\gl@align>}}
\def\gl@align#1#2{\lower.6ex\vbox{\baselineskip\z@skip\lineskip\z@
\ialign{$\m@th#1\hfil##\hfil$\crcr#2\crcr\sim\crcr}}} \makeatother

\begin{document}
\begin{flushright}
BA-08-05 \\
KEK-TH-1251
\end{flushright}
\vspace*{1.0cm}

\begin{center}
\baselineskip 20pt {\Large\bf
Higgs Boson Mass Bounds in the Standard Model \\
with Type III and Type I Seesaw }
\vspace{1cm}

{\large Ilia Gogoladze$^{a}$,
Nobuchika Okada$^{b}$
and Qaisar Shafi$^{a}$
} \vspace{.5cm}

{\baselineskip 20pt \it
$^a$Bartol Research Institute, Department of Physics and Astronomy, \\
University of Delaware, Newark, DE 19716, USA \\
\vspace{2mm} $^b$Theory Division, KEK, Tsukuba 305-0801, Japan }
\vspace{.5cm}

\vspace{1.5cm} {\bf Abstract}
\end{center}

In type III seesaw utilized to explain the observed
 solar and atmospheric neutrino oscillations
 the Standard Model (SM) particle spectrum is extended
 by introducing three SU(2)$_L$ triplet fermion fields.
This can have important implications for the SM Higgs boson mass ($M_H$)
 bounds based on vacuum stability and perturbativity arguments.
We compute the appropriate renormalization group equations
 for type III seesaw, and then proceed to identify regions
 of the parameter space such that the SM Higgs boson mass window
 is enlarged to 125 GeV $\lesssim M_H \lesssim$ 174 GeV,
 with the type III seesaw scale close to TeV.
We also display regions of the parameter space for which
 the vacuum stability and perturbativity bounds merge together
 for large neutrino Yukawa couplings.
Comparison with type I seesaw is also presented. 

\thispagestyle{empty}

\newpage

\addtocounter{page}{-1}

\baselineskip 18pt

In a recent paper \cite{HMass-typeII}
 we studied the impact of type II seesaw \cite{seesawII}
 on the predictions for the SM Higgs boson mass based
 on vacuum stability and perturbativity arguments \cite{stability1}.
It was shown in \cite{HMass-typeII} that a Higgs boson mass
 as low as 114.4 GeV (LEP2 bound \cite{LEP2})
 can be realized in type II seesaw.
This is to be contrasted with a lower bound of 131 GeV realized
 both in the absence of any seesaw or with type I seesaw \cite{seesawI}.
Thus, it would seem that discovery of
 a relatively 'light' Higgs boson (with mass well below 131 GeV)
 would signal the presence of new physics which may well be
 related to the presence of type II seesaw around the TeV scale
 or higher \cite{HMass-typeII}.
Since type II seesaw is based on SU(2)$_L$ triplet scalar fields
 it is conceivable that some of these scalars may be found at the LHC.

In this paper we wish to continue our investigation of
 the SM Higgs boson mass
 (in the light of neutrino oscillations \cite{NuData})
 by focusing our attention on type III seesaw \cite{seesawIII}.
This mechanism is very similar to type I seesaw except
 that the SM singlet fermion fields (right handed neutrinos)
 are replaced by SU(2)$_L$ triplet fermions carrying zero hypercharge.

The appearance of new fermion fields in type III seesaw modifies
 the evolution of the various SM couplings.
In particular, the renormalization group equation (RGEs)
 of the Higgs quartic coupling and the SU(2)$_L$ gauge coupling
 will be altered, and this will play an important role
 in modifying the vacuum stability ($\gtrsim$ 131 GeV)
 and perturbativity ($\lesssim$ 171 GeV)
 bounds obtained by ignoring neutrino oscillations.

The implication of type I seesaw for the Higgs boson mass
 has been investigated in Ref.~\cite{HMass-typeI} employing one-loop RGEs.
It was shown that the Higgs boson mass from the vacuum stability bound
 is pushed up as the Dirac Yukawa coupling becomes larger,
 and eventually merges with the perturbativity bound
 around $M_H \simeq 170$ GeV.
In our previous work \cite{HMass-typeII},
 we have investigated the Higgs boson mass bounds in type II seesaw.
The triplet scalar $\Delta$ in type II seesaw has both cubic
 and quartic couplings with the Higgs doublet and
 we have studied the coupled RGEs involving the Higgs
 doublet and $\Delta$.
The cubic and quartic couplings can drastically change
 the RGE evolution of the Higgs quartic coupling
 and in particular, the perturbativity bound is pushed down
 as the couplings become larger and eventually merges
 with the vacuum stability bounds.
Furthermore, for a plausible choice of parameters,
 the resultant Higgs boson mass can be as low as
 the LEP2 bound of 114.4 GeV \cite{LEP2}.

The basic structure of type III seesaw is similar to type I seesaw,
 except that instead of the singlet right-handed neutrinos,
 SU(2)$_L$ triplet fermions with zero hypercharge are introduced.
Thus, we may expect similar results for the Higgs boson mass
 bounds as in type I seesaw.
This turns out to be true for large Yukawa couplings.
However, as we will show below, the range of the Higgs boson mass
 from vacuum stability and perturbativity bounds is enlarged
 from the SM one for small Yukawa couplings.
This is in contrast with type I seesaw in which
 the Higgs boson mass bounds reduce to the SM ones
 for small Yukawa couplings.

We begin by introducing three generations of fermions
 $\psi_i$ ($i=1,2,3$),
 which transforms as $({\bf 3}, 0)$
 under the electroweak gauge group SU(2)$_L\times$U(1)$_Y$:
\bea
 \psi_i = \sum_a \frac{\sigma^a}{2} \psi_i^a
=\frac{1}{2}
 \left( \begin{array}{cc}
    \psi^0_i & \sqrt{2} \psi^{+}_i   \\
    \sqrt{2} \psi^{-}_i & - \psi^0_i   \\
 \end{array}\right) .
\eea
For simplicity, we assume in this paper that the three triplet
 fermions are degenerate in mass ($M$)
 so that their mass matrix is proportional to
 the 3 by 3 unit matrix, ${\bf M}=M \times {\bf 1}_{3 \times 3}$.

With canonically normalized kinetic terms for the triplet fermions,
 we introduce the Yukawa coupling
\bea
 {\cal L}_{Y}= y_{ij} \overline{\ell_i} \psi_j \Phi,
\eea
 where $\Phi$ is the Higgs doublet.
At low energies, the heavy triplet fermions are integrated out
 and the effective dimension five operator is generated
 by the seesaw mechanism.
After electroweak symmetry breaking, the light neutrino mass
 matrix is obtained as
\bea
  {\bf M}_\nu =
  - \frac{v^2}{8} {\bf Y}_\nu^T {\bf M}^{-1} {\bf Y}_\nu
  = - \frac{v^2}{8 M} {\bf Y}_\nu^T {\bf Y}_\nu ,
\eea
 where $v=246.2$ GeV is the VEV of the Higgs doublet,
 ${\bf Y}_\nu = y_{ij}$ is a 3$\times$3 Yukawa matrix,
 and we have used the assumption ${\bf M}=M \times {\bf 1}_{3 \times 3}$
 in the last equality.

At energies higher than the triplet fermion mass,
 the SM RGEs should be modified to include contributions
 from the triplet fermions, so that the RGE evolution
 of the Higgs quartic coupling is altered.
We have computed the one-loop RGEs for the new contributions
 associated with type III seesaw scenario.
In our analysis, we employ two-loop RGEs for the SM couplings.

For a renormalization scale $\mu < M$,
 the triplet fermions are decoupled.
For the  three SM gauge couplings, we have
\bea
 \frac{d g_i}{d \ln \mu} =
 \frac{b_i}{16 \pi^2} g_i^3 +\frac{g_i^3}{(16\pi^2)^2}
  \left( \sum_{j=1}^3 B_{ij}g_j^2 - C_i y_t^2   \right),
\label{gauge}
\eea
 where $g_i$ ($i=1,2,3$) are the SM  gauge couplings,
\bea
b_i = \left(\frac{41}{10},-\frac{19}{6},-7\right),~~~~
 { B_{ij}} =
 \left(
  \begin{array}{ccc}
  \frac{199}{50}& \frac{27}{10}&\frac{44}{5}\\
 \frac{9}{10} & \frac{35}{6}&12 \\
 \frac{11}{10}&\frac{9}{2}&-26
  \end{array}
 \right),  ~~~~
C_i=\left( \frac{17}{10}, \frac{3}{2}, 2 \right),
\label{beta} \eea {and we have included the contribution from the
top} Yukawa coupling ($y_t$). The top quark pole mass is taken to be
 the central value $M_t= 172.6$ GeV, \cite{Tevatron},
with
 $(\alpha_1, \alpha_2, \alpha_3)=(0.01681, 0.03354, 0.1176)$
 at the Z-pole ($M_Z$) \cite{PDG}.
For the top Yukawa coupling, we have \cite{RGE},
\bea \label{ty}
 \frac{d y_t}{d \ln \mu}
 = y_t  \left(
 \frac{1}{16 \pi^2} \beta_t^{(1)} + \frac{1}{(16 \pi^2)^2} \beta_t^{(2)}
 \right).
\eea
Here the one-loop contribution is
\bea
 \beta_t^{(1)} =  \frac{9}{2} y_t^2 -
  \left(
    \frac{17}{20} g_1^2 + \frac{9}{4} g_2^2 + 8 g_3^2
  \right) ,
\label{topYukawa-1}
\eea
while the two-loop contribution is given by
\bea
\beta_t^{(2)} &=&
 -12 y_t^4 +   \left(
    \frac{393}{80} g_1^2 + \frac{225}{16} g_2^2  + 36 g_3^2
   \right)  y_t^2  \nonumber \\
 &&+ \frac{1187}{600} g_1^4 - \frac{9}{20} g_1^2 g_2^2 +
  \frac{19}{15} g_1^2 g_3^2
  - \frac{23}{4}  g_2^4  + 9  g_2^2 g_3^2  - 108 g_3^4 \nonumber \\
 &&+ \frac{3}{2} \lambda^2 - 6 \lambda y_t^2 .
\label{topYukawa-2}
\eea
In solving Eq.~(\ref{ty}),
 the initial top Yukawa coupling at $\mu=M_t$
 is determined from the relation
 between the pole mass and the running Yukawa coupling
 \cite{Pole-MSbar}, \cite{Pole-MSbar2},
\bea
  M_t \simeq m_t(M_t)
 \left( 1 + \frac{4}{3} \frac{\alpha_3(M_t)}{\pi}
          + 11  \left( \frac{\alpha_3(M_t)}{\pi} \right)^2
          - \left( \frac{m_t(M_t)}{2 \pi v}  \right)^2
 \right),
\eea
 with $ y_t(M_t) = \sqrt{2} m_t(M_t)/v$, where $v=246.2$ GeV.
Here, the second and third terms in parenthesis correspond to
 one- and two-loop QCD corrections, respectively,
 while the fourth term comes from the electroweak corrections at one-loop level.
The numerical values of the third and fourth terms
 are comparable (their signs are opposite).
The electroweak corrections at two-loop level and
 the three-loop QCD corrections \cite{Pole-MSbar2},
 are of comparable and sufficiently small magnitude \cite{Pole-MSbar2}
 to be safely ignored.

The RGE  for the Higgs quartic coupling is given by \cite{RGE},
\bea
\frac{d \lambda}{d \ln \mu}
 =   \frac{1}{16 \pi^2} \beta_\lambda^{(1)}
   + \frac{1}{(16 \pi^2)^2}  \beta_\lambda^{(2)},
\label{self}
\eea
with
\bea
 \beta_\lambda^{(1)} &=& 12 \lambda^2 -
 \left(  \frac{9}{5} g_1^2+9 g_2^2  \right) \lambda
 + \frac{9}{4}  \left(
 \frac{3}{25} g_1^4 + \frac{2}{5} g_1^2 g_2^2 +g_2^4
 \right) + 12 y_t^2 \lambda  - 12 y_t^4 ,
\label{self-1}
\eea
and
\bea
  \beta_\lambda^{(2)} &=&
 -78 \lambda^3  + 18 \left( \frac{3}{5} g_1^2 + 3 g_2^2 \right) \lambda^2
 - \left( \frac{73}{8} g_2^4  - \frac{117}{20} g_1^2 g_2^2
 + \frac{2661}{100} g_1^4  \right) \lambda - 3 \lambda y_t^4
 \nonumber \\
 &&+ \frac{305}{8} g_2^6 - \frac{289}{40} g_1^2 g_2^4
 - \frac{1677}{200} g_1^4 g_2^2 - \frac{3411}{1000} g_1^6
 - 64 g_3^2 y_t^4 - \frac{16}{5} g_1^2 y_t^4
 - \frac{9}{2} g_2^4 y_t^2
 \nonumber \\
 && + 10 \lambda \left(
  \frac{17}{20} g_1^2 + \frac{9}{4} g_2^2 + 8 g_3^2 \right) y_t^2
 -\frac{3}{5} g_1^2 \left(\frac{57}{10} g_1^2 - 21 g_2^2 \right)
  y_t^2  - 72 \lambda^2 y_t^2  + 60 y_t^6.
\label{self-2}
\eea
The Higgs boson pole mass $M_H$ is determined
 by its relation to the running Higgs quartic coupling
through the one-loop matching condition \cite{HiggsPole} , \bea
 \lambda(M_H) \; v^2 = M_H^2 \left( 1+ \Delta_h(M_H) \right),
\eea
where
\bea
 \Delta_h(M_H) = \frac{G_F}{\sqrt{2}} \frac{M_Z^2}{16 \pi^2}
 \left[
   \frac{M_H^2}{M_Z^2} f_1\left(\frac{M_H^2}{M_Z^2}\right)
 + f_0\left(\frac{M_H^2}{M_Z^2}\right)
 + \frac{M_Z^2}{M_H^2} f_{-1}\left(\frac{M_H^2}{M_Z^2}\right)
 \right]
\eea
The functions are given by
\bea
f_1(\xi) &=&
   \frac{3}{2}  \ln \xi - \frac{1}{2} Z\left(\frac{1}{\xi}\right)
 -Z\left(\frac{c_w^2}{\xi}\right) -\ln c_w^2
  +\frac{9}{2} \left( \frac{25}{9} - \frac{\pi}{\sqrt{3}}
  \right), \nonumber\\
f_0(\xi)  &=& - 6\ln\frac{M_H^2}{M_Z^2}
  \left[ 1 +2 c_w^2 -2 \frac{M_t^2}{M_Z^2} \right]
 +\frac{3 c_w^2 \xi}{\xi-c_w^2} \ln \frac{\xi}{c_w^2}
  +2 Z\!\left( \frac{1}{\xi} \right) \nonumber\\
 &&+ 4 c_w^2 Z\left( \frac{c_w^2}{\xi} \right)
   +\left(\frac{3 c_w^2}{s_w^2} +12 c_w^2\right) \ln c_w^2
  -\frac{15}{2} \left( 1 +2 c_w^2 \right) \nonumber\\
 &&- 3\frac{M_t^2}{M_Z^2} \left[
     2 Z\left( \frac{M_t^2}{M_Z^2 \xi} \right)
    +4 \ln \frac{M_t^2}{M_Z^2} -5 \right], \nonumber\\
f_{-1}(\xi) &=& 6 \ln \frac{M_H^2}{M_Z^2}
  \left[ 1 +2 c_w^4 -4\frac{M_t^4}{M_Z^4} \right]
 -6 Z\left( \frac{1}{\xi} \right)-12 c_w^4 Z\left( \frac{c_w^2}{\xi} \right)
 -12 c_w^4 \ln c_w^2 \nonumber\\
 && + 8\left( 1 +2 c_w^4 \right) +24 \frac{M_t^4}{M_Z^4}
  \left[\ln \frac{M_t^2}{M_Z^2} -2 + Z\left( \frac{M_t^2}{M_Z^2 \xi} \right)
\right],
\eea
with $s_w^2 = \sin^2 \theta_W$, $ c_w^2 = \cos^2 \theta_W$
 ($\theta_W$ denotes the weak mixing angle) and
\bea
 Z(z) = \left\{
  \begin{array}{cc}
    2 A \arctan(1/A) & (z > 1/4 ) \\
    A \ln \left[ (1+A)/(1-A) \right] & (z < 1/4 ) ,
 \end{array}\right.
\eea
 with $ A = \sqrt{ \left| 1 - 4 z \right| }$.

For $ \mu \geq M$, {as previously mentioned, we include} the
one-loop contributions from the new fermion triplets. {We have
computed these new contributions for the wave function
 renormalization factors and vertex corrections
 for the neutrino Yukawa couplings, and for
 the Higgs quartic coupling,
 by employing dimensional regularization
 and $R_{\xi}$ gauge for the gauge boson propagators.}
The wave function renormalization factors of
 the SM Higgs doublet and lepton doublets receive
 new contributions from the neutrino Yukawa couplings:
\bea
 \delta Z_\Phi^{\rm new} =
 - \frac{1}{16 \pi^2 \epsilon}
 \left(
    \frac{3}{2} {\rm tr}\left[{\bf Y}_\nu^\dagger {\bf Y}_\nu \right]
  \right) , ~~~~
 \delta Z_\ell^{\rm new} =
 - \frac{1}{16 \pi^2 \epsilon}
 \left(
  \frac{3}{4} {\bf Y}_\nu {\bf Y}_\nu^\dagger \right) .
\eea
The wave function renormalization factor for
 the triplet fermions is calculated to be
\bea
 \delta Z_\psi &=&
 - \frac{1}{16 \pi^2 \epsilon}
 \left(
  \frac{1}{2} {\bf Y}_\nu^\dagger {\bf Y}_\nu
  + 4 g_2^2 \xi_2  \right),
\eea where $\xi_2$ is a gauge parameter for SU(2)$_L$. For the
neutrino Yukawa coupling matrix we find
 \bea \delta {\bf Y}_\nu =  -
\frac{1}{16 \pi^2 \epsilon}
 \left[
   \frac{3}{10} g_1^2 \xi_1
   + g_2^2 \left( 6 + \frac{7}{2} \xi_2 \right)
 \right] .
\eea The Higgs quartic coupling receives a new contribution given by
 \bea \delta \lambda =
 - \frac{1}{16 \pi^2 \epsilon}
 \left(
   \frac{5}{4} {\rm tr}
   \left[{\bf Y}_\nu^\dagger {\bf Y}_\nu \right]
  \right).
\eea
Consequently, we replace $b_i$ in Eq.~(\ref{beta}) with
\bea
  b_i = \left(\frac{41}{10}, \frac{5}{6}, -7 \right).
\eea
In the RGE for the top Yukawa coupling,
 $\beta_t^{(1)}$ is modified, {namely}
\bea
 \beta_t^{(1)} \to  \beta_t^{(1)}
 + \frac{3}{4} {\rm tr}\left[ {\bf S_\nu} \right]
\label{topYukawa-modified}
\eea
through the new contribution to the wave function
 renormalization for the Higgs doublet.
Here, ${\bf S_\nu}={\bf Y}_\nu^\dagger {\bf Y}_\nu$
 and its corresponding RGE is found  to be
\bea
 16 \pi^2 \frac{d {\bf S_\nu}}{d \ln \mu}
 = {\bf S_\nu}
  \left[
   6 y_t^2 + \frac{3}{2} {\rm tr}\left[ {\bf S_\nu} \right]
   -\left( \frac{9}{10} g_1^2 +\frac{33}{2} g_2^2 \right)
   + \frac{5}{4} {\bf S_\nu} \right] .
\label{RGE-Snu}
\eea
The RGE of the Higgs quartic coupling acquires
 a new entry in Eq.(\ref{self-1}),
\bea
 \beta_\lambda^{(1)} \to
  \beta_\lambda^{(1)}
 + 3 \; {\rm tr}[{\bf S_\nu}] \lambda
 - \frac{5}{4}{\rm tr}[{\bf S_\nu}^2] .
\label{self-1New}
\eea

We next analyze the RGEs numerically and show how
 the vacuum stability and perturbativity bounds on Higgs boson mass
 are altered in the presence of type III seesaw.
Fixing the cutoff scale as $M_{Pl}=1.2 \times 10^{19}$ GeV,
 we define the vacuum stability bound as the lowest Higgs boson mass
 obtained from the running of the Higgs quartic coupling
 which satisfies the condition $\lambda(\mu) \geq 0$
 for any scale between $M_H \leq \mu \leq M_{Pl}$.
On the other hand, the perturbativity bound is defined as
 the highest Higgs boson mass obtained from the running
 of the Higgs quartic coupling with the condition
 $\lambda(\mu) \leq \sqrt{4 \pi}$ for any scale
 between $M_H \leq \mu \leq M_{Pl}$.

In order to see the effects of the neutrino Yukawa coupling
 on the Higgs boson mass bounds, let us first examine
 a toy model with ${\bf Y}_\nu = {\rm diag}(0,0,Y_\nu)$.
In Fig.~1, the running Higgs mass, defined as
 $m_H(\mu)=\sqrt{\lambda(\mu)} v$ for the vacuum stability bound
 is depicted for various $Y_\nu$ values and a fixed seesaw scale $M=10^{13}$ GeV.
Fig.~2 shows a running Higgs mass
 for the perturbativity bound for various $Y_\nu$ values
 and a fixed seesaw scale $M=10^{13}$ GeV.
We find that as $Y_\nu$ is increased,
 the running Higgs mass for the vacuum stability bound shows
 a peak which reaches higher, while the change
 in the RGE evolution for the perturbativity bound is mild.
This means that as $Y_\nu$ is increased, the vacuum stability
 and perturbativity bounds eventually merge.
In other words, the window for the Higgs boson mass
 between the vacuum stability and perturbative bounds becomes narrower
 and is eventually closed as $Y_\nu$ becomes sufficiently large.
This behavior is shown in Fig.~3.

Note that in Fig.~3, the precise values of the Higgs boson mass bounds
 are different from the SM ones, even in the limit $Y_\nu \to 0$.
This is because the RGE of the SU(2)$_L$ gauge coupling
 for $\mu \geq M$ is different from the SM one
 in the presence of the electroweak triplet fermions.
We will see later that this effect becomes sizable
 as $M$ is lowered.

It is certainly interesting to consider more realistic cases
 so as to reproduce the current neutrino oscillation data.
The light neutrino mass matrix is diagonalized
 by a mixing matrix $U_{MNS}$ such that
\bea
  {\bf M}_\nu = \frac{v^2}{8 M} \; {\bf S}_\nu
   = U_{MNS} D_\nu U^T_{MNS}
\label{Mix}
\eea
with $D_\nu ={\rm diag}(m_1, m_2, m_3)$,
 where we have assumed, for simplicity, that
 the Yukawa matrix ${\bf Y}_\nu$ is real.
We further assume that the mixing matrix has
 the so-called tri-bimaximal form \cite{hps},
\bea
U_{MNS}=
\left(
\begin{array}{ccc}
\sqrt{\frac{2}{3}} & \sqrt{\frac{1}{3}} & 0 \\
-\sqrt{\frac{1}{6}} & \sqrt{\frac{1}{3}} &  \sqrt{\frac{1}{2}} \\
-\sqrt{\frac{1}{6}} & \sqrt{\frac{1}{3}} & -\sqrt{\frac{1}{2}}
\end{array}
\right) ,
\label{ansatz}
\eea
 which is in very good agreement with the current
 best fit values of the neutrino oscillation data \cite{NuData}.
Let us consider two examples for the light neutrino mass spectrum,
 the hierarchical case and the inverted-hierarchical case.
In the hierarchical case, we have
\bea
 D_\nu \simeq
 {\rm diag}(0,\sqrt{\Delta m_{12}^2}, \sqrt{\Delta m_{23}^2}),
\eea
while for the inverted-hierarchical case, we choose
\bea
 D_\nu \simeq
 {\rm diag}(\sqrt{-\Delta m_{12}^2 + \Delta m_{23}^2},
 \sqrt{\Delta m_{23}^2}, 0).
\eea
 We fix the input values for the solar and atmospheric
 neutrino oscillation data as \cite{NuData}
\bea
 \Delta m_{12}^2 &=& 8.2 \times 10^{-5} \; {\rm eV}^2 \nonumber \\
 \Delta m_{23}^2 &=& 2.4 \times 10^{-3} \; {\rm eV}^2.
 \label{massdiff}
\eea

From Eqs.~(\ref{Mix})-(\ref{massdiff}),
 we obtain the matrix ${\bf S_\nu}$ as a function of $M$,
 with which we numerically solve the RGEs and obtain
 the Higgs boson mass bounds as a function of $M$.
The window for the Higgs boson pole mass for both
 the hierarchical and inverted-hierarchical cases
 is shown in Fig.~4.
As $M$, or equivalently the Yukawa couplings become large,
 the window for the Higgs boson mass becomes narrower
 and is eventually closed.

As previously mentioned, for low $M$ values,
 the Higgs boson mass bounds with type III seesaw are different from the SM ones
 and the range of the Higgs boson mass window is enlarged.
This result can be qualitatively understood in the following way.
The presence of the triplet fermions significantly alters
 the RGE running of the SU(2)$_L$ gauge coupling by making it
 asymptotically non-free, so that $g_2(\mu)$ for $\mu > M$
 is larger than the SM value without type III seesaw.
In the analysis of the vacuum  stability bound,
 the Higgs quartic coupling is small so that Eq.~(\ref{self-1})
 can be approximated as
\bea
 \beta_\lambda^{(1)} \simeq
  \frac{9}{4}  \left(
 \frac{3}{25} g_1^4 + \frac{2}{5} g_1^2 g_2^2 +g_2^4
 \right) - 12 y_t^4.
\eea
The first term on the right hand side is larger
 in type III seesaw than in the SM case and as a result,
 the Higgs quartic coupling decreases more slowly than in the SM.
Consequently, the vacuum stability bound on the Higgs boson mass is
lowered. For the perturbativity bound, the Higgs quartic coupling is
large and
 Eq.~(\ref{self-1}) can be approximated by
\bea
 \beta_\lambda^{(1)} \simeq 12 \lambda^2 -
 \left(  \frac{9}{5} g_1^2+9 g_2^2  \right) \lambda
 + 12 y_t^2 \lambda  - 12 y_t^4,
\eea
The beta function is smaller than the SM one due to the second term.
Therefore, the evolution of the Higgs quartic coupling is slower,
 and as a result, the Higgs boson mass based on the perturbative bound
 is somewhat larger than the SM one.

Finally, it is interesting to compare the results in type III seesaw
 with those in type I seesaw \cite{HMass-typeI}.
We examine type I seesaw with three singlet fermions
 with a degenerate mass $M$.
The RGE formulas for type I seesaw are slightly different from
 those in type III.
The RGEs for the gauge couplings are the same
 as in the SM because only SM singlet fields are introduced.
The RGEs corresponding to Eqs.~(\ref{topYukawa-modified})-(\ref{self-1New})
 are given by
\bea
&& \beta_t^{(1)} \to  \beta_t^{(1)}
 + {\rm tr}\left[ {\bf S_\nu} \right], \nonumber \\
&&  16 \pi^2 \frac{d {\bf S_\nu}}{d \ln \mu}
 = {\bf S_\nu}
  \left[
   6 y_t^2 + 2 \; {\rm tr}\left[ {\bf S_\nu} \right]
   -\left( \frac{9}{10} g_1^2 +\frac{9}{2} g_2^2 \right)
   + 3 \; {\bf S_\nu} \right] , \nonumber \\
&&  \beta_\lambda^{(1)} \to
  \beta_\lambda^{(1)}
 + 4 \; {\rm tr}[{\bf S_\nu}] \lambda
 - 4 \; {\rm tr}[{\bf S_\nu}^2] .
\label{RGE-typeI}
\eea
In type I seesaw, we define the light neutrino mass matrix
 as ${\bf M}_\nu = - \frac{v^2}{2 M} {\bf Y}_\nu^T {\bf Y}_\nu $,
 as usual.

We repeat the previous analysis with the RGEs for the type I seesaw
 and the results are shown in Fig.~5, to be compared with
 the results for type III shown in the Fig.~4.
We find qualitatively the same behavior for the Higgs boson mass bounds
 for a large seesaw scale or equivalently large Yukawa couplings.
An important difference between type I and III seesaws
 can be seen for a small seesaw scale or equivalently small Yukawa couplings.
In type I seesaw, the Higgs boson mass bounds reduce to the SM results,
 because only the Yukawa couplings affect the RGEs in this case.

In conclusion, we have considered the potential impact of
 type III seesaw on the vacuum stability and perturbativity bounds
 on the Higgs boson mass.
For energies higher than the seesaw scale, the triplet fermions
 introduced in type III seesaw are involved in quantum corrections
 and the RGEs of the SM  are modified.
There are two important effects.
One is the neutrino Yukawa coupling contribution and the other
 is the modification of the RGE of the SU(2)$_L$ gauge coupling
 due to the presence of the triplet fermions.
We have found that as the Yukawa couplings are increased,
 the vacuum stability bound grows and eventually merges
 with the perturbativity bound.
Therefore, the Higgs boson mass window is closed at some large
 Yukawa couplings with a fixed seesaw scale, or some high seesaw
 scale by fixing the light neutrino  mass scale.
Even if the new Yukawa couplings are negligible,
 there is a remarkable effect due to the modification
 of the RGE evolution of the SU(2)$_L$ gauge coupling.
For a low seesaw scale, the Higgs boson mass window
 between the vacuum stability and perturbative bounds
 turns out to be wider than the SM one.
This is in contrast with type I seesaw
 where the Higgs boson mass bounds in the SM are reproduced
 in the small Yukawa coupling limit.

\section*{Acknowledgments}
This work is supported in part by
 the DOE Grant \# DE-FG02-91ER40626 (I.G. and Q.S.),
 and the Grant-in-Aid for Scientific Research from the Ministry
 of Education, Science and Culture of Japan,
 \#18740170 (N.O.).


\newpage
\begin{figure}[t]
\includegraphics[scale=1.2]{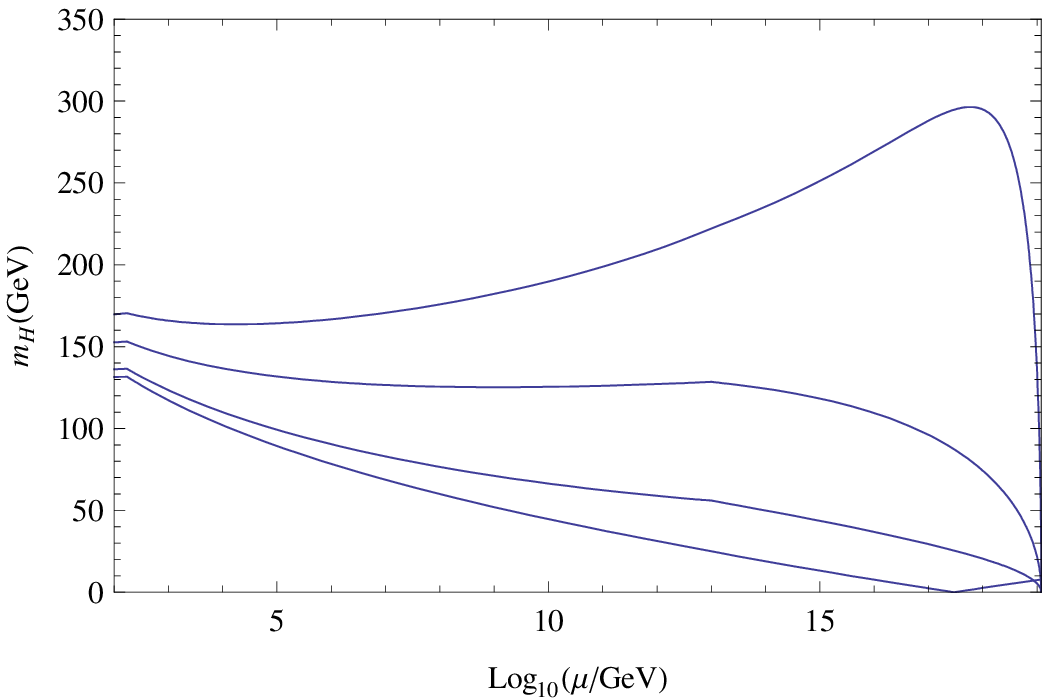}
\caption{
Evolution of running Higgs mass
($m_H(\mu)=\sqrt{\lambda(\mu)} v$)
 corresponding to the vacuum stability bound
 for various $Y_\nu$ values and the seesaw scale $M=10^{13}$ GeV.
Each line corresponds to
 $Y_\nu=$1.5, 1.2, 0.8 and 0 from top to bottom.
}
\end{figure}
\begin{figure}[t]
\includegraphics[scale=1.2]{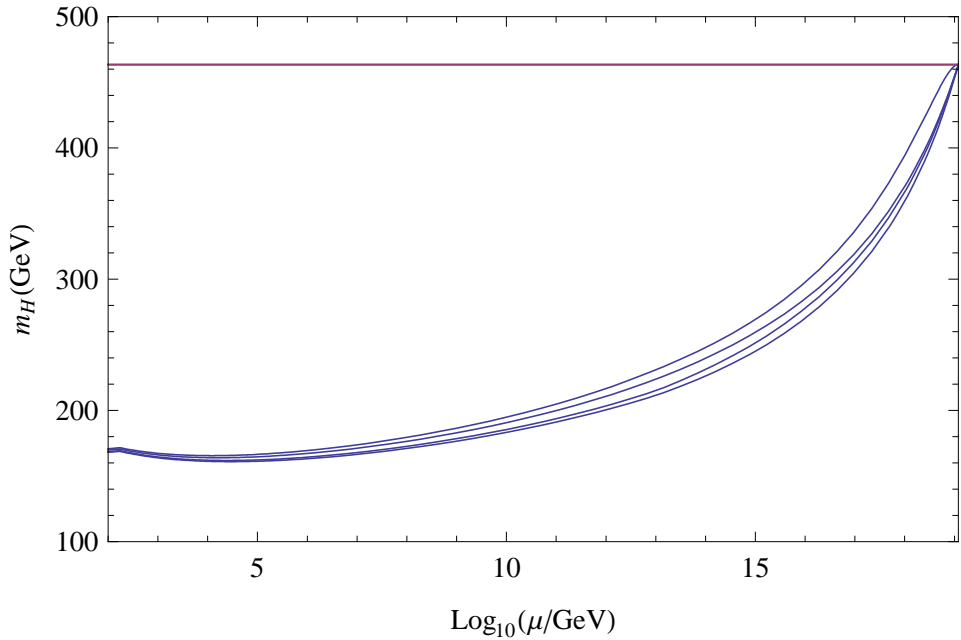}
\caption{
Evolution of running Higgs mass
($m_H(\mu)=\sqrt{\lambda(\mu)} v$)
 corresponding to the perturbativity bound
 for various $Y_\nu$ values and the seesaw scale $M=10^{13}$ GeV.
Each line corresponds to
 $Y_\nu=$1.5, 1.2, 0.8 and 0 from top to bottom.
The horizontal line corresponds to
 $m_H(M_{Pl})=(4 \pi)^{1/4} v= 464$ GeV.
}
\end{figure}
\begin{figure}[t]
\includegraphics[scale=1.2]{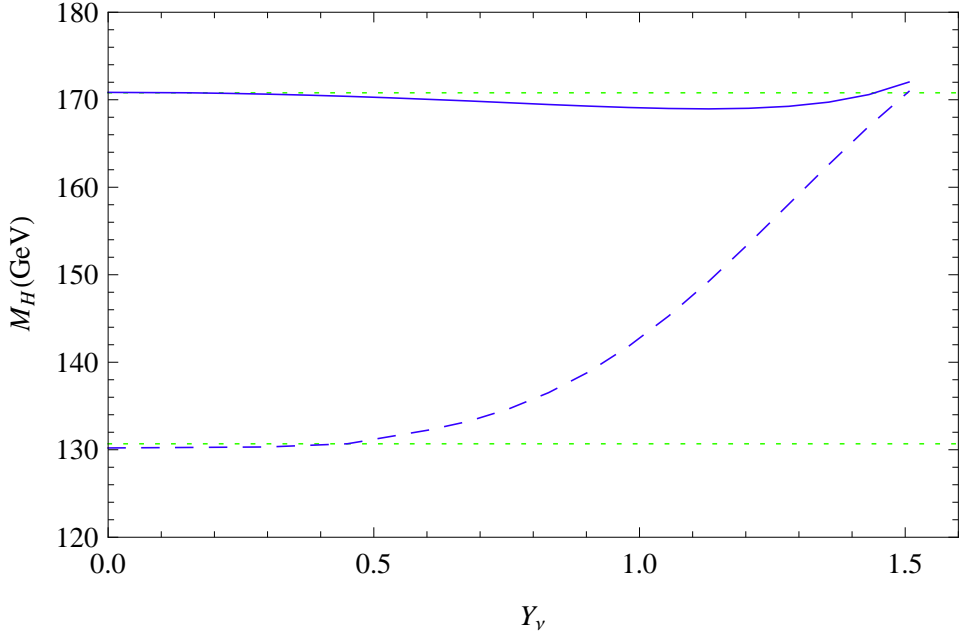}
\caption{ Perturbativity (solid) and vacuum stability (dashed)
 bounds on the Higgs boson pole mass ($M_H$) versus
 $Y_\nu$ with the seesaw scale $M=10^{13}$ GeV.
The upper and lower dotted lines respectively show
 the perturbativity bound ($M_H\simeq$171 GeV)
 and the vacuum stability bound ($M_H\simeq$131 GeV)
 in the SM case.
}
\end{figure}
\begin{figure}[t]
\includegraphics[scale=1.2]{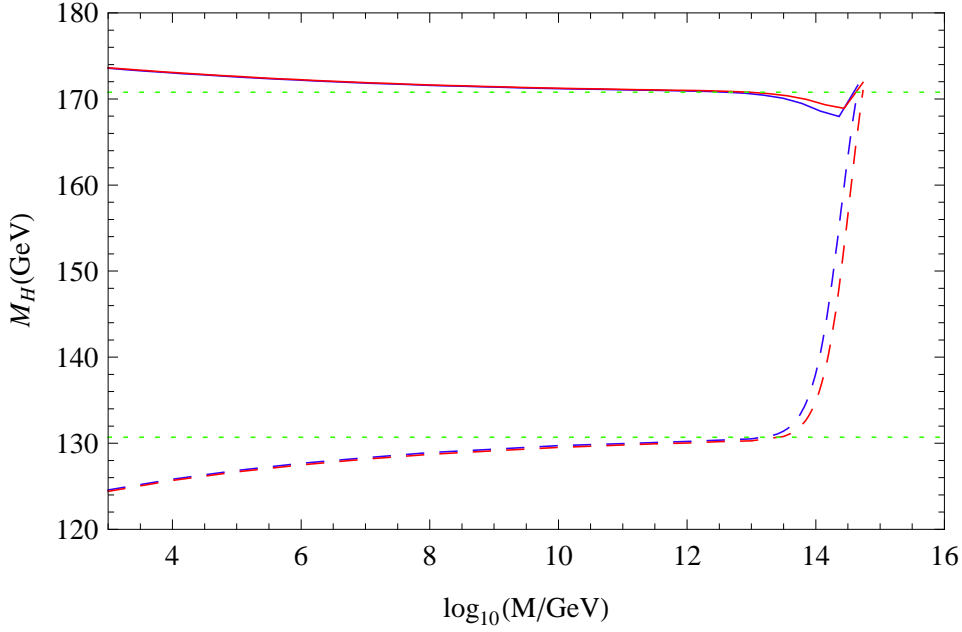}
\caption{ Perturbativity and vacuum stability bounds versus
 $M$, with a hierarchical mass spectrum
 (outer region in red), and an inverted-hierarchical
 mass spectrum (inner region in blue).
}
\end{figure}
\begin{figure}[t]
\includegraphics[scale=1.2]{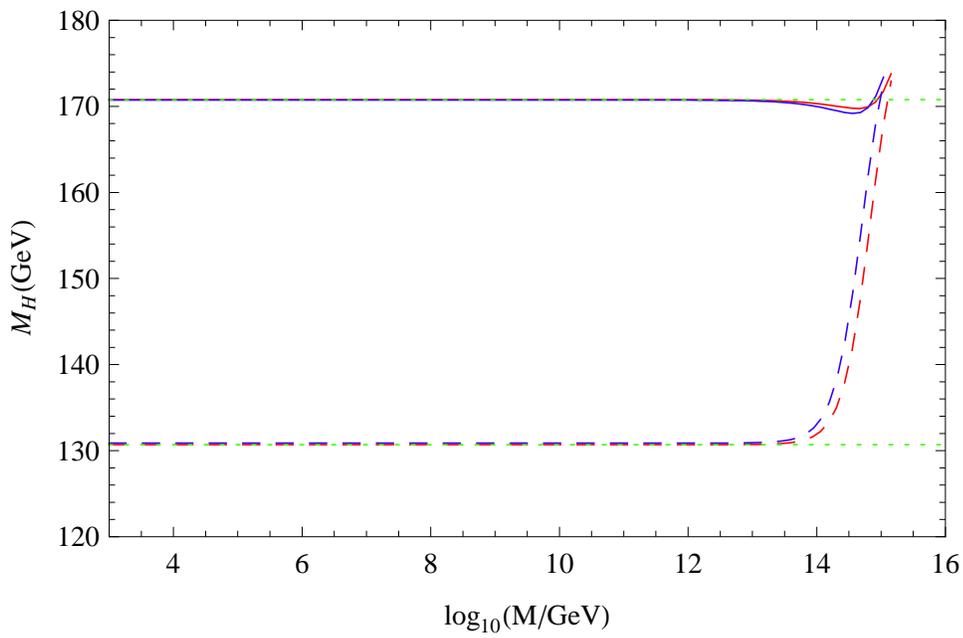}
\caption{ Similar to Figure 4, but for type I seesaw with three
singlet neutrinos. }
\end{figure}

\end{document}